# No space-time singularity in black-hole physics

José Bernabéu
Department of Theoretical Physics, University of Valencia, and
IFIC (CSIC-UV), 46100 Burjassot, Valencia, Spain

Massive stellar gravitational collapse is not an endless process. The Standard Model of particle physics predicts the existence of a repulsive interaction, two- neutrino mediated, with a coherent weak charge of macroscopic matter proportional to the number of constituents. After gravity, this is the second longest range -microns- force and, below nanometers, its magnitude overcomes the attractive gravitation. The layer distribution of pressure inside a compact neutronized uniform sphere, independent of its size, is compared for the two interactions with opposite pressure gradient. Taking as reference the mean square radius, the volume evolution of the pressure, for fixed mass and weak charge, leads to an equilibrium scenario with a characteristic black-hole radius $R_B$ = 1.58 nm. This size scale should be universal, independent of the mass, as long as the mass is nearly proportional to the weak charge.

The existence of static solutions of General Relativity with a space-time singularity was first predicted [1] by Schwarzschild in 1916 for a spherical massive object. In addition, there is also a coordinate singularity at a radius proportional to the mass

$$r_s = 2G\frac{M}{c^2} \tag{1}$$

called the Schwarzschild radius, where G is the gravitational constant, M is the object mass and c is the speed of light. It is interpreted as an event- horizon. The concept of black-hole was born as an object with radius smaller than its Schwarzschild radius. In fact, the Schwarzschild radius can be obtained in the Newtonian limit as the distance in which the escape velocity equals the speed of light. The event-horizon emits Hawking radiation [2].

The existence of black-holes in nature came as a consequence of the understanding of the physics of stellar evolution for star masses above 3 solar masses. The normal life of stars is kept by the equilibrium condition between the attractive gravitational pressure and the repulsive pressure of thermonuclear fusion reactions until the formation of the most stable Fe, Co, Ni nuclides. For stars around 2 solar masses the subsequent collapse and the bounce of the supernova explosion leaves a remnant of a compact neutron star in equilibrium with the degeneracy pressure for neutrons. Typical nuclear densities and sizes of tens of Km. are then the appropriate scales.

However, for more massive stars, the neutron degeneracy pressure cannot compensate the gravitational pressure and the star continues its collapse into a black hole. At present massive star black holes have been observed by a variety of probes as well as black holes from other origins, like the supermassive black holes at the center of galaxies. Our Sagitarius A* in the Milky Way has about one million of solar masses. The natural language of star collapse will be used here.

The question discussed in this letter is whether one expects in nature that the collapse would continue without end, leading to the existence of the space-time singularity at the center of the black hole. The answer in general relativity, the source being the energy density of the object, is well known and given by Penrose's singularity theorem [3]: "The formation of a trapped surface implies that space-time contains a singularity". The theorem assumes that matter has a non-negative energy density It guarantees that geodesic incompleteness occurs inside any black-hole whenever matter satisfies these energy conditions. Possible violations of this assumption have been discussed by Zeldovich [4] at very high densities and by Candelas [5] in a quantum context. These studies assume that the dynamical evolution of matter is entirely governed within the geometrical realm of general relativity where the matter trajectory is given by the geodesics of the curved space-time. It takes for granted that no other effect in nature, outside gravitation, will be able to compete with the gravitational collapse at these high densities.

In this letter, a non-speculative novel mechanism is discussed providing a new Action associated to an existing interaction. It is generated by an internal coherent source -proportional to density- not related to space-time properties. The standard model of particle physics predicts that, between neutral baryonic matter, there is a weak neutral current interaction mediated by two neutrinos -and so of long range behaviour below microns- with all appropriate properties to reestablish an equilibrium scenario with the gravitational action. The gauge character of the interaction guarantees a repulsive effect, contrary to attractive gravitation. The weak neutral current coherent source is proportional to the number of constituents. For an extended object it is then proportional to the weak charge density. The new Action implies that the matter dynamics departs from the geodesic evolution dictated by general relativity. Starting from the weak neutral current interaction between two neutronized baryons of weak charges $N_1$ and $N_2$, the aim of this letter is to obtain the effective force-field inside an extended compact object of baryonic matter of weak charge N and the corresponding pressure gradient. This is compared with the gravitational pressure gradient inside this object of mass M. Their different volume dependences then leads to a definite universal equilibrium.

The singularity result is taking for granted that there is no other Action in nature able to compete with the gravitational effects in the collapse dynamics. On the contrary, it is shown below that, in agreement with the Standard Model of particle physics, there is a weak neutral current interaction of baryonic matter with all appropriate properties to re-establish an equilibrium scenario with the gravitational action. The gravitational source is coherent with the total energy and, for extended objects, the gravitational effects depend on the energy density. For neutral (of electric charge) aggregate matter there exists a long-range repulsive weak force with the essential property of a coherent weak charge proportional to the number of constituents. The Standard Model of particle physics predicts such interaction associated with the two neutrino mediation. For extended objects, this weak interaction becomes proportional to the weak charge density.

In the context of weak charged currents for electrons, the interaction mediated by two massless electron-neutrinos was first obtained in 1968 by Feinberg and Sucher [6]. It was later extended to include neutral current interactions for any system using both the dispersion theory approach and time-ordered perturbation theory [7, 8, 9]. Recently [10] the long range weak effective potential between neutral aggregate matter was obtained including all ingredients on neutrino properties known from neutrino oscillations [11, 12, 13, 14], like masses and mixings. The aim is to find, in this $\Delta L = 0$ effect, a regime of virtual non-relativistic neutrinos providing a sensitivity of the interaction to the absolute neutrino mass, distinguishing its either Dirac or Majorana nature. A conceptual design for its observation with atomic clock interferometry has been presented [15].

The $2\nu$ potential is obtained [10] from the integral transform of the t-channel absorptive part of the amplitude as depicted in Fig. 1

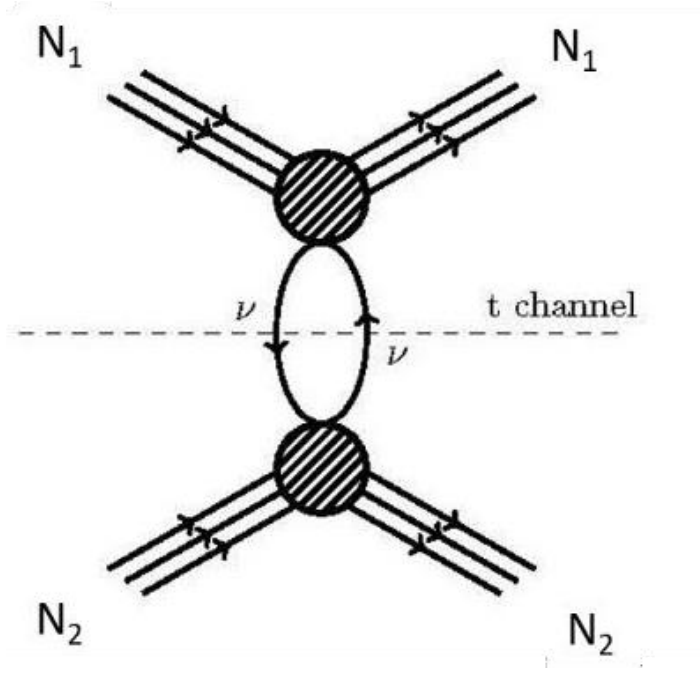

Fig. 1. The longest range weak interaction amplitude mediated by two neutrinos for aggregate matter.

The expected range of the interaction is in the scale of microns and there its magnitude is weaker than the gravitational interaction by about 12 orders of magnitude. However, at distances below the micron scale the neutrino mass is inoperative and so it is the flavor mixing too. In this regime the weak interaction behaves with distance as an inverse power law for all pairs of neutrino exchanges. Neutral aggregate matter has a coherent weak charge proportional to the number of constituents. For the case of neutronized matter, the interaction is only due to the weak neutral current charges and so diagonal and universal for all neutrino species. As a consequence, the interaction of two baryons of weak charges $N_1$ and $N_2$ is the sum of products of diagonal charges of the matter, i.e., $[Q_1].\{Q_2\}= 3\ N_1\ N_2$. The potential energy under these conditions is thus given by

$$\mathbb{V}_w = G_F^2\ \frac{3\ N_1 N_2}{16\pi^3}\frac{1}{r^5}; \quad \forall (m_i r) \ll 1 \qquad (2)$$

in natural units $\hbar = c = 1$, with $G_F$ the Fermi coupling of weak interactions and $m_i$ the neutrino masses. As a consequence, at distances of nanometers (nm) one expects that this repulsive weak interaction will compete with the attractive gravitation and its effects have to be considered. It is worth to emphasize again that we are not talking about a possible exotic force but on an actual force present in the Standard model of particle physics. After gravitation, this is the second longest range interaction existing in nature for the neutral matter.

For the black-hole study, our initial system is a neutronized object of definite mass $M$ - including internal energies - and weak charge $N$ with the approximate relation $M = m\ N$, $m$ being the neutron mass, in gravitational collapse until it reaches the micron scale - the range of the weak force -. Under the influence of this new Action the collapse dynamics will be altered, not controlled by the geodesics dictated by general relativity. For the extended object with total weak charge $N$, the inner interaction becomes dependent on the weak charge density and thus independent of the actual nature of the constituent degrees of freedom at high densities. Gravitation and this weak neutral current force are the only long range density-dependent -energy density versus weak charge density-interactions relevant to volume effects for compact neutral massive objects. The calculation of the volume dependence, at fixed $M$ and $N$, of the two pressures is now undertaken.

In general relativity, the way of discussing equilibrium for (spherically symmetric, cold) objects is through the so-called TOV (Tolman-Oppenheimer-Volkoff) equations [16]. The object is modeled by a perfect fluid with energy density and pressure (measured in the local rest frame). The basic equation provides the pressure gradient and it has its Newtonian limit which is going to be used for the gravitational action. With the advent of the weak interaction pressure, a treatment of equilibrium as the balance between these two pressures is most appropriate. A sphere of uniform density with radius R is taken. The pressure depends on the inner layer at distance $r < R$, so we have to compare the two force-fields, per mass unit in gravitation and per weak charge unit in the two-neutrino interaction. Its calculation in the last case presents some subtleties. The potential (2) is very singular at short distances and not valid below hundreds of fermis, where shorter range interactions mediated by other exchanged massive particles enter. This regulator acts as a physical cut-off $r_c$ at short distances, with the validity of Eq. (2) extended to distances between microns and $r_c$ of order of hundreds of fermis, much below nanometers.

- **The inner force-field**

Eq. (2) is the potential energy in vacuum between the two weak charges $N_1$ and $N_2$. We consider now the potential field -per unit of weak charge- $U_w(r)$ created by an extended neutral baryon of radius R and total weak charge N as the source. At the point of the probe $\vec{r}$ one has contributions from the variable volume element sources around $\vec{r}'$. The integration over the relative polar angle of $\vec{r}'$ with $\vec{r}$ then leads to

$$U_w = G_F^2 \frac{3N}{16\pi^3} \frac{1}{2R^3 r} \cdot \int_0^R dr' r' \left\{ \frac{1}{|r - r'|^3} - \frac{1}{(r + r')^3} \right\} \quad (3)$$

For r >> R, Eq. (3) reproduces Eq. (2) for $N_1 = N$ and $N_2 = 1$. As expected, in the internal region r < R the first term of Eq. (3) depends on the physical short distance cut-off $r_c$ discussed above. However, this dependence is a constant independent of $r$, not affecting the force field per particle $w(r)$ given by

$$w(r < R) = G_F^2 \frac{3N}{16\pi^3} \frac{r}{R^3} \frac{5R^2 - r^2}{(R^2 - r^2)^3} \quad (4)$$

This result for the weak force field per weak charge unit should be compared in its r-dependence and sign with the well known gravitational force field per mass unit g(r) given by

$$g(r < R) = -G \frac{M}{R^3} r \quad (5)$$

with opposite sign to that of the weak force-field w(r < R).

- **The layer distribution of pressure**

We start reminding the results for the gravitational pressure and then obtain the corresponding distribution for the weak pressure.

For an extended system of given density, **the force per volume unit gives the pressure gradient** dP/dr, where P is the pressure at each layer identified by x = r/R , ∀ x ∈ [0,1) , independent of the size of the object.

From Eq. (5) the gravitational pressure gradient is obtained by the product of the force-field with the mass density. With the boundary condition $P_g(R, x = 1) = 0$, the gravitational layer distribution is

$$P_g(R, x) = G \frac{3}{5} \frac{M^2}{4\pi R^4} f_g(x);$$

$$f_g(x) = \frac{5}{2}(1 - x^2) \quad (6)$$

where the volume dependence of $P_g$ at fixed *M* and its *x*-distribution have been factorized by the introduction of the $f_g(x)$ function. The factorization has been made by taking the mean square radius as the reference layer $\overline{x^2} = \frac{3}{5}$ . The function $f_g(x)$ shows the negative pressure gradient of gravitation.

From Eq. (4), the weak pressure gradient is obtained by the product of the weak force-field with the weak charge density. With the boundary condition $P_w(R, x = 0) = 0$, the weak layer distribution is

$$P_w(R, x) = \frac{G_F^2}{16\pi^3} \frac{27N^2}{2\pi R^8} f_w(x);$$

$$f_w(x) = \frac{x^2(5 - 3x^2)}{12(1 - x^2)^2} \quad (7)$$

where the volume dependence of $P_w$ at fixed *N* and its *x*-distribution have been factorized by using again the same layer at the mean square radius as reference. The function $f_w(x)$ shows the positive pressure gradient of the weak force. It is compared with $f_g(x)$ in Fig. 2

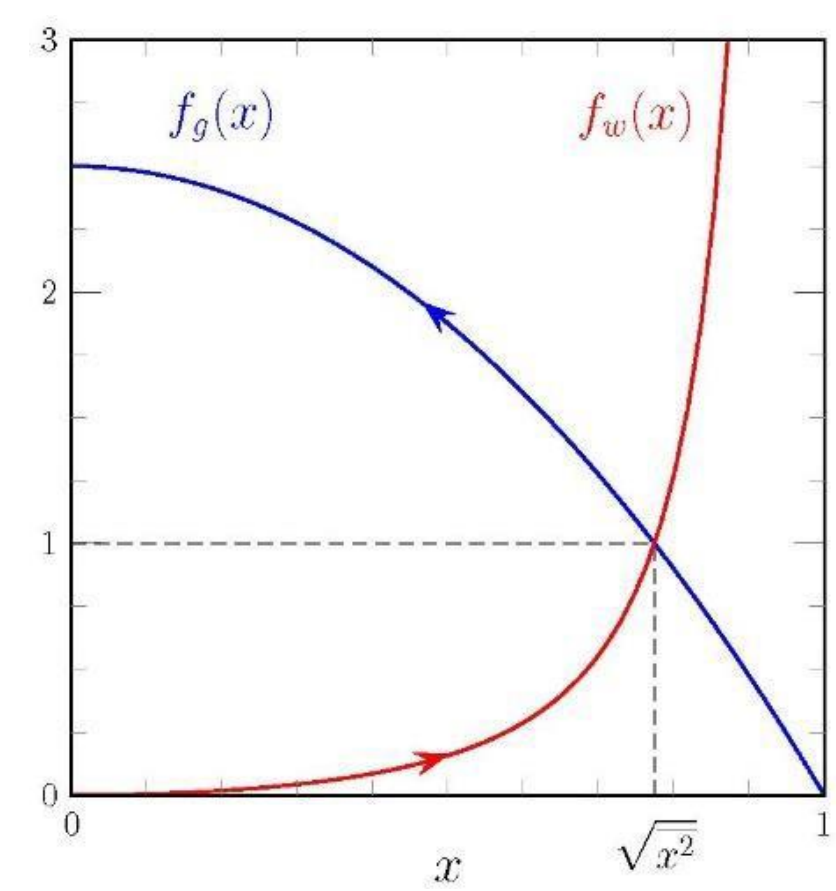

Fig. 2. The layer distributions of the gravitational and weak pressures $f_g(x)$ –blue- and $f_w(x)$ -red-. The arrows indicate the sense of the pressure gradient.

- **The volume dependence at fixed M and N**

As already mentioned, the volume dependence of the pressures for the two interactions $P_g$, with fixed *M*, and $P_w$, with fixed *N*, is compared for the reference layer in which $f_g = f_w = 1$. The comparison of Eqs. (6) and (7) shows a different behavior in the size, as $V^{-4/3}$ and $V^{-8/3}$ respectively, with V the volume.

This fact and the opposite gradient lead to an equilibrium scenario with a characteristic size of the black-hole defined by the radius

$$R_B = \left[\frac{G_F^2}{G}\frac{5}{2\pi^3}\left(\frac{3N}{2M}\right)^2\right]^{\frac{1}{4}} \quad (8)$$

The assumed equality of the distributions for the mass and weak charge densities leads to an equilibrium size of the object, Eq. (8), in terms of the global mass M and weak charge N. The near proportionality between M and N gives a result independent of the mass with the value $R_B = 1.58$ nm. Consistently, this is well within the region of distances in which the weak potential, Eq. (2), is valid.

Needless to say, the internal structure of this matter under these conditions is either described in the realm of high density QCD or simply unknown. Nevertheless, as already indicated, only the global properties of the system, with the coherent weak charge and mass - including internal energy-, are relevant to find the equilibrium condition between the pressures from the two density-dependent interactions in the extended object. These properties were established when the system became neutronized.

To conclude, the reasoning developed in this letter shows that the existence of the long range two-neutrino mediated weak neutral current interaction has the properties to give a novel insight into the problem of massive gravitational collapse. The long-range weak interaction is repulsive, with a coherent weak charge for neutral aggregate matter and having a spatial dependence other than gravity. The calculation of the inner force-field due to this interaction predicts its pressure balance with that of the gravitational effect. The black-holes can reach an equilibrium size with a characteristic radius $R_B$ =1.58 nm, quite independent of the black-hole mass. The result of this novel mechanism for the resolution of the singularity problem motivates the study of the combined dynamics of the two interactions in order to find the approach to equilibrium of this massive neutral object.

**Acknowledgements**

The author acknowledges comments and suggestions from G. Barenboim, F. Botella, C. Espinoza, D. Espriu, J. Navarro, M. Nebot, S. Palomares and A. Santamaria. This research has been supported by the Spanish AEI-MICINN, PID2020 - 113334GB - I00/AEI/ 10.13039501100011033 and by the Generalitat Valenciana grant CIPROM/2021/054.